\documentclass[aps,prl,twocolumn,showpacs,superscriptaddress]{revtex4}
\usepackage{bbm}
\usepackage{mathrsfs}
\usepackage{epsfig}
\usepackage{graphicx}
\usepackage{amsfonts}
\usepackage[figuresright]{rotating}
\usepackage{amssymb}
\usepackage{amsmath}
\usepackage{dcolumn}
\usepackage{bm}

\def\avg#1{\langle#1\rangle}

\def\be{\begin{equation}} \def\ee{\end{equation}}
\def\bea{\begin{eqnarray}} \def\eea{\end{eqnarray}}

\def\nn{\nonumber}

\begin{document}
\title{Pomeranchuk cooling of the SU($2N$) ultra-cold fermions in
optical lattices}
\author{Zi Cai}
\affiliation{Department of Physics, University of California, San
Diego, CA92093}
\affiliation{Department of Physics and Arnold
Sommerfeld Center for Theoretical Physics,
Ludwig-Maximilians-Universit{\"a}t M{\"u}nchen, Theresienstr.\ 37,
80333 Munich, Germany}
\author{Hsiang-hsuan Hung}
\affiliation{Department of Physics, University of California, San
Diego, CA92093} \affiliation{Department of Electrical and Computer
Engineering, University of Illinois, Urbana-Champaign,  Illinois
61801}
\author{Lei Wang}
\affiliation{Theoretische Physik, ETH Zurich, 8093 Zurich,
Switzerland}
\author{Dong Zheng}
\affiliation{Department of Physics, Tsinghua University, Beijing, China 100084}
\author{Congjun Wu}
\affiliation{Department of Physics, University of California, San
Diego, CA92093}
\affiliation{Key Laboratory of Artificial Micro- and Nano-structures
at the Ministry of Education, the School of Physics and Technology
at Wuhan University, Wuhan 430072, China
}

\begin{abstract}
We investigate the thermodynamic properties of a half-filled
SU($2N$) Hubbard model in the two-dimensional square lattice by the
method of the determinant quantum Monte Carlo simulation, which is
free of the fermion ``sign problem''. The large number of
hyperfine-spin components enhances spin fluctuations, which
facilitates the Pomeranchuk cooling to temperatures comparable to
the superexchange energy scale in the case of SU$(6)$. Various
physical quantities including entropy, charge fluctuations, and spin
correlations are calculated.
\end{abstract}
\maketitle

The SU($2N$) and Sp($2N$) symmetries are usually studied in high
energy physics. They were introduced to condensed matter physics
originally as a mathematic convenience. For example, large-$N$
analysis was performed for the SU($2N$) symmetric Heisenberg models
to systematically handle strong correlation effects
\cite{arovas1988,affleck1988,read1989, read1991}, while  realistic
electron systems are usually only SU(2) invariant. However, with the
recent development of the ultra-cold atom physics, fermion systems
with SU($2N$) and Sp($2N$) symmetries are not just of purely
academic interests, but are currently under experimental
investigations. It was first pointed out in Ref. \cite{wu2003} that
large spin alkali and alkaline-earth fermion systems can exhibit
these high symmetries. For example a generic Sp(4), or,
isomorphically SO(5) symmetry, can be realized for fermion systems
with the hyperfine spin $F=\frac{3}{2}$ without fine-tuning
\cite{wu2003,wu2006a}. This Sp(4) symmetry can be further augmented
to SU(4) for alkaline-earth fermions, such as $^{135}$Ba,
$^{137}$Ba, and $^{201}$Hg  because their interactions are
hyperfine-spin independent \cite{wu2003}. Experimentally, both the
fermionic atoms of $^{173}$Yb and $^{87}$Sr have been cooled down to
quantum degeneracy \cite{taie2010,desalvo2010,hideaki2011}. The
$^{173}$Yb ($F=I=\frac{5}{2}$) and $^{87}$Sr ($F=I=\frac{9}{2}$)
systems exhibit the SU(6) and SU(10) symmetries, respectively. Using
alkaline-earth fermions to study the SU$(2N)$ symmetry was also
proposed in Ref. [\onlinecite{gorshkov2010}].

The SU($2N$) Hubbard model exhibits interesting phenomena that are
absent in the standard SU(2) formulation. It is known that quantum
spin fluctuations are enhanced by the large number of fermion
components \cite{wu2010a}. This effect gives rise to exotic quantum
magnetism in large-spin ultra-cold fermi systems with high
symmetries
\cite{wu2005a,cazalilla2009,manmana2011,hung2011,assaad2005,Szirmai2011a,
Szirmai2011b,Vekua2011}. For example, various SU($2N$) valence-bond
solid and spin liquid states have been proposed that have not been
observed in solid state systems before \cite{chen2005,
hung2011,hermele2011}. In addition, as we will show below, the
multi-component nature of the SU($2N$) Hubbard model also
significantly lowers the charge gap of the Mott-insulating states at
the intermediate interaction strengths comparable to the bandwidth.

In this paper, we focus on the temperature regime ($t>T \sim J$),
which is of current experimental interest. Here $t$ denotes the
hopping integral of the Hubbard model, $J=4t^2/U$ is the
antiferromagnetic exchange energy scale, and $U$ is the onsite
repulsion. The thermodynamic properties of the half-filled SU($2N$)
Hubbard model in the 2D square lattice are studied by  determinant
quantum Monte-Carlo (DQMC) simulations
\cite{Scalapino2006,Hirsch1985}, which is an unbiased,
non-perturbative method. It is free of the sign problem at
half-filling, thus high numerical precision can be achieved down to
low temperatures ($T/t \sim 0.1$). (Recently, the high temperature
properties of the SU($2N$) Hubbard model has been studied from
series expansions, which are only accurate at $T\gg \max(t, U)$
\cite{hazzard2010}.) Special attentions are devoted to the
interaction-induced adiabatic coolings. We find that the system can
be cooled down to the temperature scale at $J$ from an initial
temperature accessible in current experiments. This Pomeranchuk
cooling effect, though it is very weak in the SU$(2)$ Hubbard model
\cite{paiva2010,dare2007}, is enhanced in the SU$(6)$ case.

We consider the following SU($2N$) Hubbard model defined in
the 2D square lattice at half-filling as
\bea
H=-t\sum_{\avg{i,j},\alpha}\big(c^\dag_{i\alpha}c_{j\alpha}+h.c.\big)
+\frac{U}{2}\sum_i\big(n_i-N \big)^2,
\label{Eq:sun}
\eea
where $\alpha$ runs over the $2N$ components; $\avg{i,j}$ denotes the
summation over the nearest neighbors; $n_i$ is the total particle
number operator on site $i$ defined as $n_i=\sum_{\alpha=1}^{2N}
c^\dag_{i\alpha}c_{i\alpha}$.
The chemical potential $\mu$ is set to 0 and thus does not appear explicitly.
Eq. \ref{Eq:sun} is invariant under
the particle-hole transformation in bipartite lattices. Similarly to
the case of SU$(2)$, it is easy to prove that the sign problem is
also absent for the half-filled SU$(2N)$ Hubbard model of Eq.
\ref{Eq:sun} in bipartite lattices in the DQMC simulations.

Below we will present our DQMC simulations of thermodynamic
quantities of the SU$(2N)$ Hubbard model with $2N=4$ and $6$ on a $L
\times L$ square lattice with the periodical boundary condition. The
second order Suzuki-Trotter decomposition is used. The Trotter steps
are taken to be $\Delta \tau=\beta/M$, where $\beta=1/T$ is the
inverse of the temperature $T$ and $M$ ranges from $30$ to $150$
depending on temperatures. We have checked that the simulation
results converge with varying the values of $\Delta \tau$. Instead
of using the Hubbard-Stratonovich (HS) transformation in the spin
channel \cite{Hirsch1983}, we adopt the HS transformation in the
charge channel which maintains the $SU(2N)$ symmetry explicitly
\cite{Assaad1998}. This method gives rise to errors on the order of
$(\Delta \tau)^4$.

Before presenting numerical results, let us explain qualitatively
how the SU($2N$) generalization of the Hubbard model makes their
charge and magnetic properties different from those of the SU(2)
case. When deeply inside the Mott insulating state, magnetic
properties at low temperatures are determined by superexchange
processes. The number of superexchange processes between a pair of
nearest-neighbor sites in the SU($2N$) case scales as $N^2$. This
means that the SU($2N$) generalization enhances magnetic quantum
fluctuations, and thus weakens, or even completely suppresses the
long-range antiferromagnetic (AF) correlations. These strong
magnetic fluctuations greatly enhance the entropy in the the
temperature regime ($U>T>J$), which is high enough to suppress
short-range AF correlations but not sufficient to unfreeze charge
fluctuations.

\begin{figure}[htb]
\includegraphics[width=0.45\linewidth]{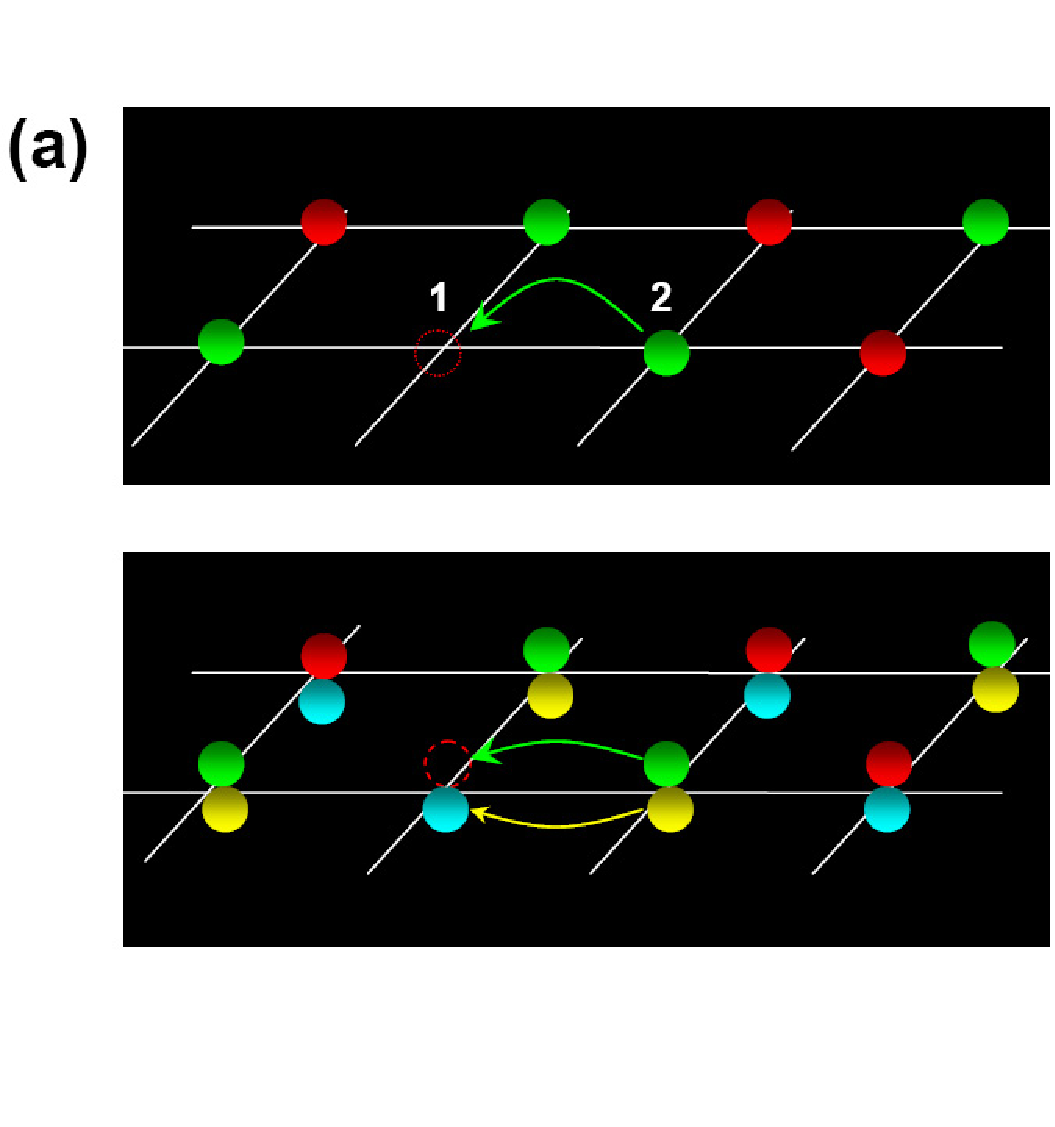}
\includegraphics[width=0.53\linewidth]{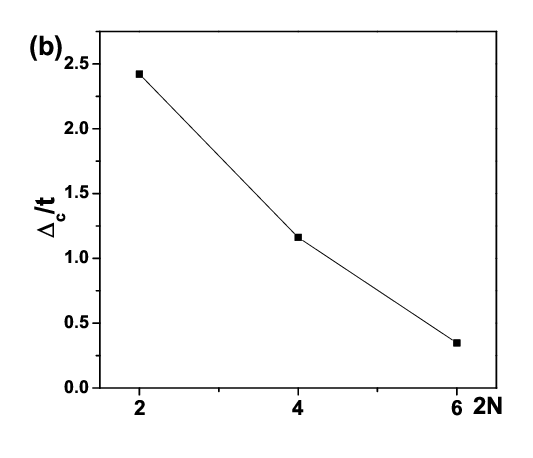}
\caption{(a) Sketches of a hole hopping in the SU(2) (up) and SU(4)
(down) AF backgrounds, respectively.
(b) Charge gaps as a function of $2N$ at $L=10$ and $U/t=8$.
}
\label{fig:hole}
\end{figure}

The charge properties in the Mott insulating state are characterized
by  the charge gap $\Delta_c$: the energy cost to add a particle or
a hole into the system. The half-filling case is a particle-hole
symmetric point, and thus a particle or hole excitation each cost
the same energy for the grand canonical Hamiltonian Eq.
\ref{Eq:sun}. In the atomic limit ($U/t\rightarrow \infty$), the
charge gap is $\Delta_c\rightarrow \frac{U}{2}$,  which is
independent of $2N$. However, for the intermediate interactions
comparable with the bandwidth, propagations of the extra particle
(hole) in the AF background can significantly lower the charge gap.
In Fig. \ref{fig:hole} (a), we compare the hopping of an extra hole
in the AF background of the half-filled SU(2) and SU(4) Mott
insulators. In the SU(4) case, there is more than one way for the
hole to hop from one site to another. The mobility of the extra hole
is increased and thus, in the SU($2N$) Mott insulating state, the
charge gap is much lower compared to the SU(2) case. We perform the
zero temperature projector QMC to extract the charge gap  from the
unequal-time single-particle correlation functions (see the
supplementary material) as shown in Fig. \ref{fig:hole} (b), which
verifies the above argument. Though the charge gap is a ground state
property, it is closely related to the thermodynamic properties and
Pomeranchuk cooling in the temperature regime we will study
($J<T<U$). Below we will show that the differences of the magnetic
and charge properties between the SU($2N$) and SU(2) cases
facilitate the Pomeranchuk cooling.

Now we address the possibility of cooling down the system by
adiabatically increasing interactions. For spinful fermion systems
(e.g. $^3$He), the Pomeranchuk effect refers to the fact that
increasing temperatures can lead to solidification because the
entropy (per particle) in the localized solid phase is larger than
that of the itinerant Fermi liquid phase. The reason is that, in the
Fermi liquid phase, only fermions close to Fermi surfaces within $T$
contribute to entropy. In solids  however,  each site contributes to
nearly $\ln 2\approx 0.69$ if $T$ is comparable to the spin exchange
energy scale of $J$, which is much smaller than the Fermi energy. In
the lattice systems near or at half-filling, increasing interactions
suppresses charge fluctuations and drives systems to the
Mott-insulating state, thus we would expect Pomeranchuk cooling
while adiabatically increasing interactions
\cite{werner2005,Li2011}. However, the situations are complicated by
the AF spin correlations which lift the huge spin degeneracy and
reduce the entropy in the Mott-insulating state. Actually, for the
SU(2) Hubbard model, both at 2D and 3D, DQMC simulations show that
the effect of Pomeranchuk cooling is not obvious with interactions
up to $U/t\sim 10$ \cite{dare2007,paiva2011,paiva2010}.

\begin{figure}[htb]
\includegraphics[width= \linewidth]{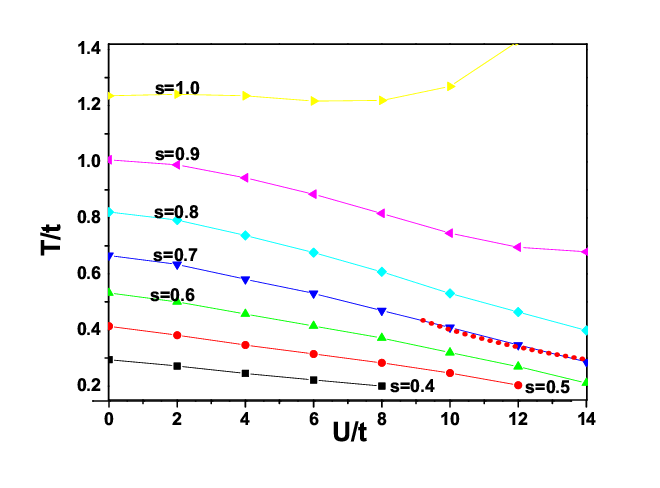}
\caption{The isoentropy curves for the half-filled SU(6) Hubbard
model on a 10$\times$10 square lattice. The dashed line denotes the
spin superexchange scale in strong coupling regime $J/t=4t/U$.}
\label{fig:isoentropy}
\end{figure}

To investigate the different behaviors between the SU($2$) and
SU($2N$) (-say, $2N=6$) fermions during the Pomeranchuk cooling, we
compare the
 ``entropy capability" (average entropy per atoms) for the half-filled SU(2)
and SU(6) Mott insulating states at the same temperature $T$ and
$U$. We focus on the temperature regime ($J<T<U$). For a certain
$T$, the entropy of the Mott insulating state comes from two
channels: the spin channel dominated by the spin degeneracy, and the
charge channel determined by excitations above the charge gap. As we
analyzed above, the AF correlations, which lift the spin degeneracy
in the SU(2) case, are weakened by the SU($2N$) symmetry. Thus the
entropy from the spin contribution in a half-filled SU(6) Mott
insulator is larger than that of the SU(2) case. This indicates that
the SU(6) Mott insulators have more ``entropy capacity" than the
SU(2) ones. For example, in the single-atom limit, the spin
entropies per atom for SU(2) and SU(6) saturate to the values of
$S_{SU(2)}=\ln[C_2^1]=0.693$ and $S_{SU(6)}=\ln[C_6^3]/3=0.998$,
respectively, for temperature $T\gg J$. Considering the charge
channel will further strengthen this tendency. Since the charge gap
of the SU(6) Mott insulating state is smaller than that of the SU(2)
case at the same value of $U$, it is easier to create excitations
above the charge gap in the SU(6) case, which further increases
entropy. The larger entropic capability of the SU(6) Mott insulating
state indicates that it is easier to exhibit the Pomeranchuk effect.

We have confirmed the above picture by performing DQMC simulations.
The entropies of the SU(6) Hubbard model are calculated for various
parameter values of $T$ and $U$, and the isoentropy curves are
displayed in Fig. \ref{fig:isoentropy}. The simulated entropy per
particle (not per site) is defined as $S_{su(2N)}=S/(N L^2)$, where
$S$ is the total entropy in the lattice. It is calculated from the
formula \bea \frac{S_{su(2N)}(T)}{k_B}=\ln 4+\frac{E(T)}T-
\int^\infty_T dT' \frac{E(T')}{T'^2}, \eea where $\ln 4$ is the
entropy at the infinite temperature or, equivalently, $T\gg U$;
$E(T)$ denotes the average internal energy per particle at
temperature $T$. For low values of the entropy, adiabatically
increasing $U$ leads to a significant cooling to a temperature
comparable to the magnetic superexchange scale $J$, which is an
important goal in current cold atom experiments. This is of direct
relevancy to the current experimental progress in ultracold
$^{173}$Yb atoms \cite{taie2010,Taie2012}.

\begin{figure}[htb]
\includegraphics[width=0.89\linewidth,height=0.6\linewidth]{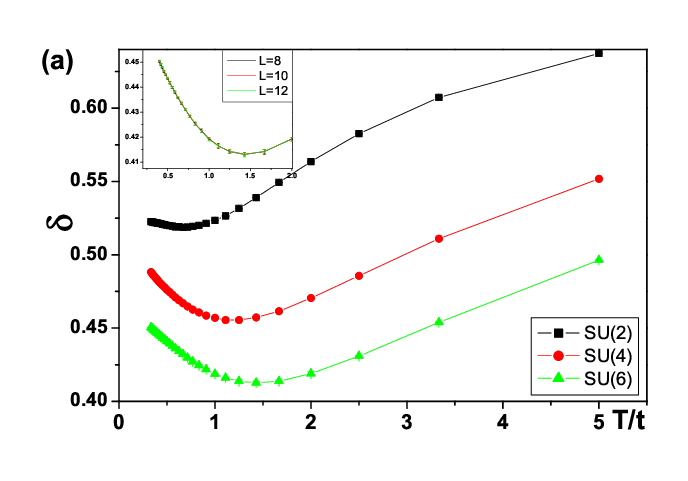}
\includegraphics[width=0.89\linewidth]{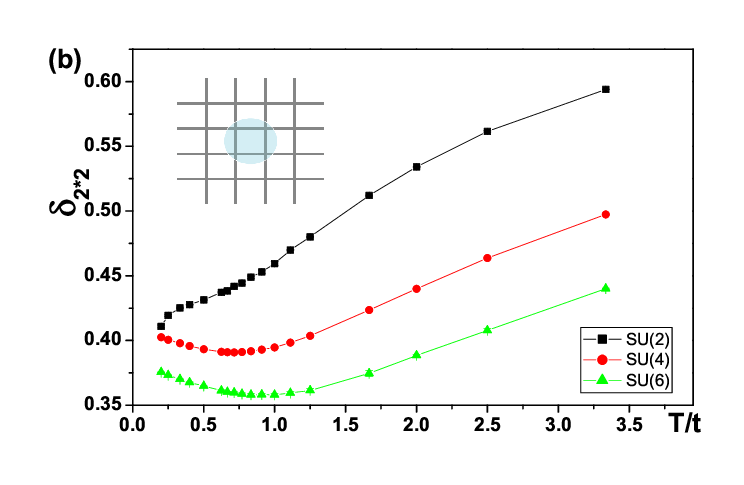}
\caption{Particle number fluctuations v.s. $T$ with parameters
$U/t=4$ and different values of $2N$ on a $10\times 10$ lattice. a)
The on-site density fluctuations $\delta_{su(2N)}(T)$. The inset
shows the convergence of $\delta_{su(6)}(T)$ with $L=8,10$, and $12$
for the SU(6) case. b) The local particle number fluctuations in a
$2\times 2$ sub-volume.  } \label{fig:CF}
\end{figure}

Next we study particle number fluctuations for the half-filled
SU($2N$) Hubbard model. The normalized on-site particle fluctuations
are defined as \bea \delta_{su(2N)}=\sqrt{\frac{\langle
n_i^2\rangle-\langle n_i\rangle^2}N}, \label{Eq:Charge} \eea where
$\langle n_i \rangle=N$. At $T\rightarrow \infty$, $\delta_{su(2N)}$
can be calculated exactly. It is independent of $2N$ as
$\delta_{su(2N)} (T\rightarrow \infty)=\frac{\sqrt 2}{2}\approx
0.71$, which acts as an upper bound on the fluctuations. Similarly,
at $U=0$, $\delta_{su(2N)}=\frac{\sqrt 2}{2}$ and is independent of
both $2N$ and $T$. For the general case, we plot the DQMC simulation
results of $\delta$ at a relatively weak interaction strength of
$U/t=4$ over a large range of temperatures seen in Fig. \ref{fig:CF}
(a). For all the cases, $\delta_{su(2N)}$ is suppressed by $U$ away
from the upper limit of $\frac{\sqrt 2}{2}$. For the cases of
SU$(4)$ and SU$(6)$, $\delta_{su(2N)}(T)$ first falls as $T$
increases, which is a reminiscence of the Pomeranchuk effect. Then,
after reaching a minimum at $T$ comparable to $t$, $\delta_{su(2N)}$
grows with increasing $T$. This indicates that fermions are
localized most strongly at an intermediate temperature scale at
which the spin channel contribution to entropy dominates. In
comparison, the non-monotonic behavior of $\delta_{su(2N)}$ is weak
in the SU(2) case. The above data agrees with the picture that large
values of $2N$ enhance spin fluctuations and thus the Pomeranchuk
effect. We also calculate the local particle number fluctuations in
a small sub-volume: $\delta_{sub}=\sqrt{[\langle
\hat{n}_{sub}^2\rangle-\langle \hat{n}_{sub}\rangle^2]/\langle
\hat{n}_{sub}\rangle}$ ($\hat n_{sub}$ is the total particle number
operator within the sub-volume).
As shown in Fig. \ref{fig:CF} (b), for the SU($4$) and SU($6$)
cases, the local density fluctuations in a $2\times 2$ sublattice
$\delta_{2\times2}$ also exhibit non-monotonic behavior similarly to
the case of the on-site density fluctuation.


\begin{figure}[htb]
\includegraphics[width=0.49\linewidth]{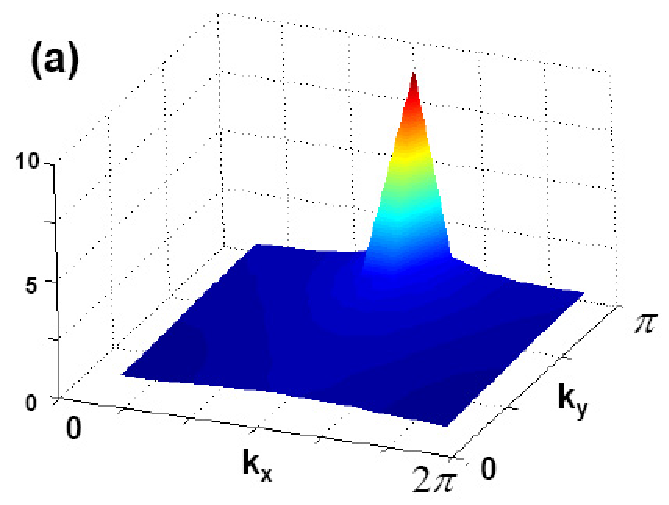}
\includegraphics[width=0.48\linewidth]{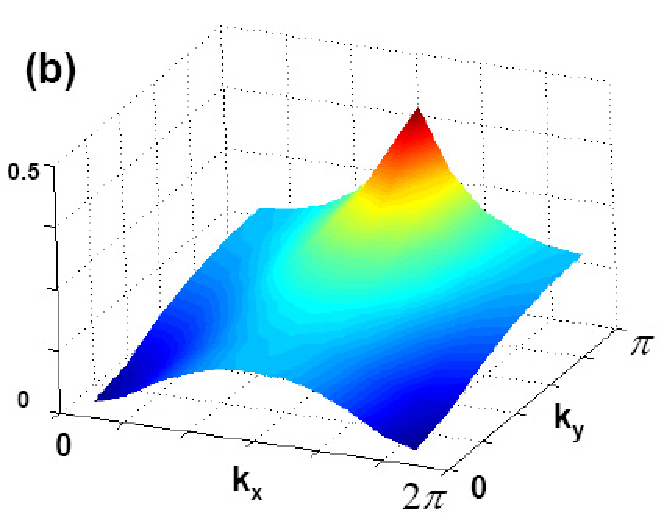}
\includegraphics[width=0.48\linewidth]{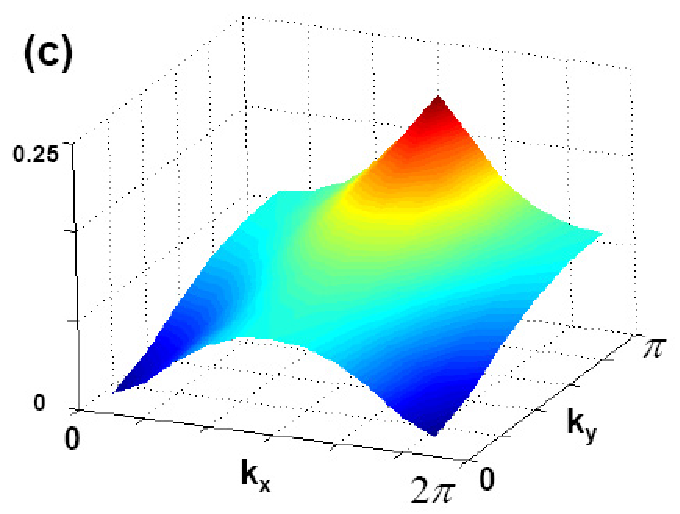}
\caption{The normalized spin structure factor $S(\vec{q})$
for the half-filled SU($2N$) Hubbard models with $2N$ equal to
(a) 2, (b) 4, and (c) 6. Parameter values are $T/t=0.1$, and $U/t=8$.
}
\label{fig:structure}
\end{figure}

Next we study spin correlations of the SU($2N$) Hubbard model. The
SU($2N$) generators can represented through fermion operators
$c_{i,\alpha} (\alpha=1 \sim 2N)$ as
$S_{\alpha\beta,i}=c^\dag_{\alpha,i}c_{\beta,i}-\frac{1}{2N}
\delta_{\alpha\beta}n_i$. There are only $(2N)^2-1$ independent
operators due to the constraint $\sum_{\alpha} S_{\alpha\alpha}=0$.
They satisfy the commutation relations  $[S_{\alpha\beta,i},
S_{\gamma\delta,j}]=
\delta_{i,j}(S_{\alpha\delta,i}\delta_{\gamma\beta}
-S_{\gamma\beta,i}\delta_{\alpha\delta})$. We define the SU($2N$)
version of the two-point equal-time spin-spin correlation as \bea
S_{spin}(i,j)=\frac{1}{(2N)^2-1} \sum_{\alpha,\beta}\langle
S_{\alpha\beta,i} S_{\beta\alpha,j}\rangle. \label{SS} \eea The spin
structure factors $S_{su(2N)}(\vec q)$ are calculated at
half-filling and a low temperature, which are defined as \bea
S_{su(2N)}(\vec{q})=\frac{1}{NL^2}\sum_{\vec{i},\vec{j}} e^{i\vec{q}
\cdot \vec r} M_{spin}(i,j), \eea where $\vec r$ is the relative
vector between sites $i$ and $j$. The distributions of
$S_{su(2N)}(\vec q)$ with $2N=2,4,6$ are plotted in Fig.
\ref{fig:structure} a), b) and c), respectively. The sharpness of
the peaks at $\mathbf{q}=(\pi,\pi)$ indicates the dominant AF
correlations in all the cases. With increasing $2N$, peaks are
broadened showing a weakening of the AF correlations.

The current experimental limit to the entropy per particle for the
two-component systems is $S_{su(2)}\sim 0.77k_B$. The corresponding
temperature scale is $T\sim t$, which is still larger than $J$
\cite{Jordens2010}. In contrast, as we analyzed above, the SU($6$)
Mott insulating state has more ``entropy capacity", which means that
for a fixed entropy per atom, the corresponding temperature of the
half-filled SU($6$) Mott insulating state is lower than that of the
SU(2) case. As shown in Fig. \ref{fig:isoentropy}, for
$S_{su(6)}\sim 0.77k_B$, the corresponding temperature of the Mott
insulating state ($U/t=12$) can reach the border of the magnetic
superexchange scale $J$. As for the experimental consequences of the
Pomeranchuk cooling, though it is difficult to directly measure
temperatures in the lattice, the non-monotonic behavior of the local
particle fluctuations, shown in Fig. \ref{fig:CF} (a) and (b), can
be tested by high-resolution {\it in situ} measurements which have
been used to observe the antibundching in ultracold atom Fermi gases
\cite{muller2010}. Repeated measurements of the local particle
numbers of identically prepared systems give rise to particle
fluctuations within the observed volume, which may contain one or
several lattice sites. Recently, the Pomeranchuk cooling has been
observed in $^{173}$Yb fermions in optical lattice (SU(6) Hubbard
model). However, we should point  an important difference between
the experiment and our calculation, namely that the filling factor
in the experiment \cite{Taie2012} is $1/6$ (one fermion per site) as
suppose to the assumed half-filling in our simulations.

In conclusion, we have performed  DQMC simulation for the
thermodynamic properties of the 2D SU($2N$) Hubbard model at
half-filling in the temperature regime of direct interest to current
experiments. The large numbers of fermion components enhance spin
fluctuations, which facilitates the Pomeranchuk cooling to
temperatures comparable to the superexchange energy scale. We have
focused on half-filling, though it is interesting to ask whether the
Pomeranchuk cooling can appear in other filling factors, especially
in the the case of 1/6-filling corresponding to one atom per site in
the SU(6) model. In this case, DQMC is plagued by the sign problem.
Though in some situations, for example at high temperatures or small
values of $U$, DQMC can still give rise to reliable results if the
sign problem is not severe.


We acknowledge J. Hirsch , D. Greif , R. Scalettar and Y. Takahashi
for helpful discussions. This work was supported by the NSF
DMR-1105945 and the AFOSR FA9550- 11-1- 0067(YIP) program.

\newpage

\begin{appendix}
\section{Supplementary material }
\author{Zi Cai}
\affiliation{Department of Physics, University of California, San
Diego, CA92093} \affiliation{Department of Physics and Arnold
Sommerfeld Center for Theoretical Physics,
Ludwig-Maximilians-Universit{\"a}t M{\"u}nchen, Theresienstr.\ 37,
80333 Munich, Germany}
\author{Hsiang-hsuan Hung}
\affiliation{Department of Physics, University of California, San
Diego, CA92093} \affiliation{Department of Electrical and Computer
Engineering, University of Illinois, Urbana-Champaign,  Illinois
61801}
\author{Lei Wang}
\affiliation{Theoretische Physik, ETH Zurich, 8093 Zurich,
Switzerland}
\author{Dong Zheng}
\affiliation{Department of Physics, Tsinghua University, Beijing,
China 100084}
\author{Congjun Wu}
\affiliation{Department of Physics, University of California, San
Diego, CA92093} \affiliation{Key Laboratory of Artificial Micro- and
Nano-structures at the Ministry of Education, the School of Physics
and Technology at Wuhan University, Wuhan 430072, China } \maketitle

In this supplementary material, we investigate thermodynamic
quantities including compressibility and nearest-neighbor spin-spin
correlations. These quantities, though not directly related with the
Pomeranchuk cooling, are of direct interests in current experiments
in ultracold atom physics. They provide a comprehensive
understanding of thermodynamical properties of the SU($2N$) Hubbard
model at half-filling.

\begin{figure}[htb]
\includegraphics[width=0.8\linewidth,height=0.6\linewidth]{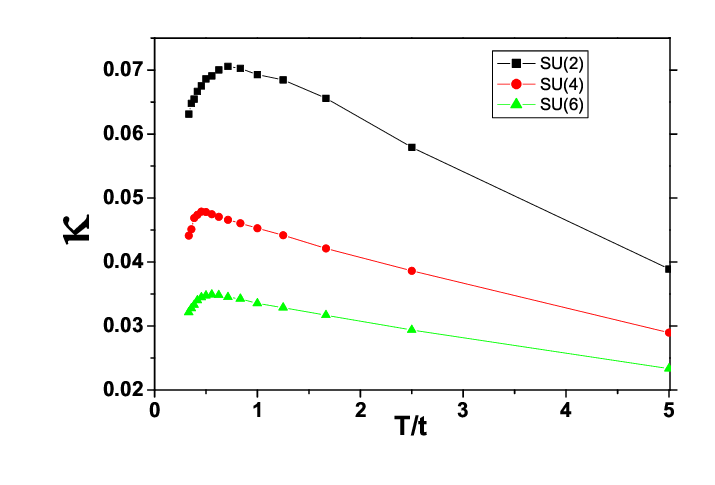}
\caption{ The normalized compressibility $\kappa_{su(2N)}/(2N)$ v.s.
$T$ at $U/t=4$ for $2N=2,4$, and 6.} \label{fig:kappa}
\end{figure}

{\it Compressibility} The compressibility $\kappa$ can be expressed
in terms of the global charge fluctuations as \bea
\kappa_{su(2N)}=\frac{1}{L^2}\frac{\partial N_f}{\partial
\mu}=\frac{1}{T L^2}(\langle \hat{N}_f^2\rangle-\langle
\hat{N}_f\rangle^2), \eea where $\hat N_f=\sum_i \hat n_i$ is the
total fermion number operator in the lattice; $\mu$ is the chemical
potential. In Fig. \ref{fig:kappa}, we plot the simulated results
for the normalized $\kappa_{su(2N)}/N$, i.e., the contribution to
$\kappa_{su(2N)}$ per fermion component. They behave similarly to
each other. $\kappa_{su(2N)}$ scales as $1/T$ like ideal gas at high
temperatures, while they are suppressed at low temperatures. At zero
temperature, $\kappa_{su(2N)}$ is suppressed to zero due to the
charge gap in the Mott-insulating states. $\kappa_{su(2N)}$ reaches
the maximum at an intermediate temperature scale which can be
attributed to the energy scale of charge fluctuations.

\begin{figure}[htb]
\includegraphics[width=0.8\linewidth,height=0.6\linewidth]{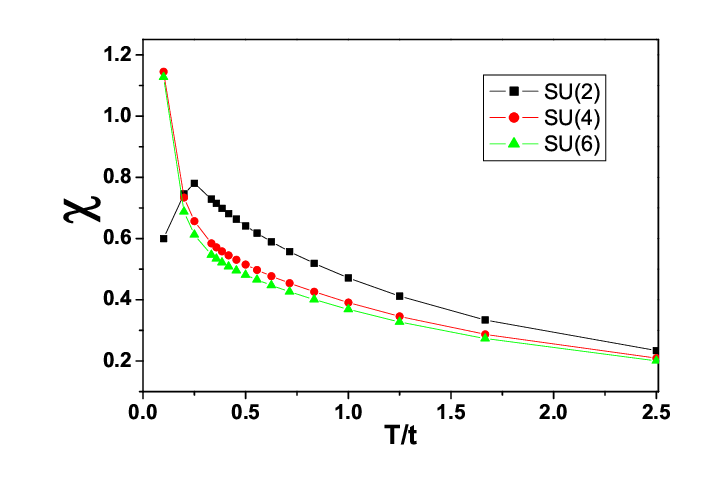}
\caption{The normalized SU($2N$) susceptibilities $\chi_{su(2N)}$
v.s. $T$ with fixed $U/t=4$ for $2N=2,4$, and $6$.}
\label{fig:spin-sus}
\end{figure}

{\it Spin susceptibility} At finite temperatures, no magnetic
long-range-order should exist in the 2D half-filled SU($2N$) model
due to its continuous symmetry. The normalized uniform SU($2N$) spin
susceptibility is defined as \bea
\chi_{su(2N)}(T)=\frac{\beta}{NL^2}\sum_{\vec{i},\vec{j}}
M_{spin}(i,j). \eea The DQMC simulation results are presented in
Fig. \ref{fig:spin-sus} for $U/t=4$. At high temperatures,
$\chi_{su(2N)}$ exhibits the standard Curie-Weiss law which scales
proportional to $1/T$. $\chi_{su(2)}$ reaches the maximum at an
intermediate temperature at the scale of $J$ below which
$\chi_{su(2)}$ is suppressed by the AF exchange. At the lowest
temperature we simulated, we did not observe the suppressions of
$\chi_{su(2N)}$ for $2N=4$ and 6. The nature of the ground states of
half-filled $SU(2N)$ Hubbard model remains an open question in
literatures when $2N$ is small but larger than $2$. Nevertheless, we
expect that they are either AF long-range-ordered like the case of
$SU(2)$, or quantum paramagnetic with or without spin gap like in
the large-$N$ limit. In either case, $\chi_{su(2N)}$ should be
suppressed to zero with approaching zero temperature.

\begin{figure}[htb]
\includegraphics[width=0.8\linewidth,height=0.6\linewidth]{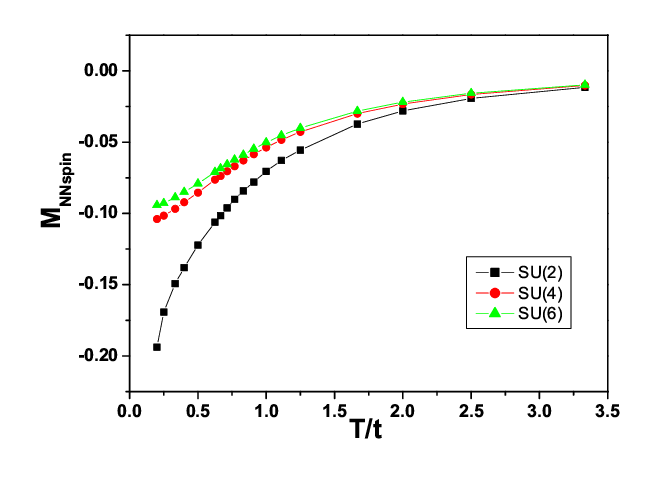}
\caption{The normalized NN spin-spin correlation v.s. $T/t$ at
$U/t=4$ for $2N=2,4$, and 6.} \label{fig:NNspin}
\end{figure}

{\it The nearest-neighbor spin-spin correlation} The
nearest-neighbor (NN) spin-spin correlation in the SU($2N$) Hubbard
model is defined as: \bea
 M_{spinNN}=\frac{1}{(2N)^2-1}
\sum_{\alpha,\beta} \langle S_{\alpha\beta,i}
S_{\beta\alpha,j}\rangle, \eea where $i$ and $j$ are two
nearest-neighbor lattice sites, and
$S_{\alpha\beta,i}=c^\dag_{\alpha,i}c_{\beta,i}-\frac{1}{2N}
\delta_{\alpha\beta}n_i$. For the SU(2) Hubbard model, the NN
correlations have been probed recently using the lattice modulation
technique \cite{Greif2011}. The NN spin-spin correlations v.s. $T/t$
for fixed $U/t$ and different $2N$ have been plotted in Fig.
\ref{fig:NNspin}. Notice that the monotonic behavior of NN spin-spin
correlations as a function of $T$ indicates that these quantities
can be used to measure temperatures and entropy in the
Mott-insulating states.

\begin{figure}[htb]
\includegraphics[width=0.8\linewidth,height=0.6\linewidth]{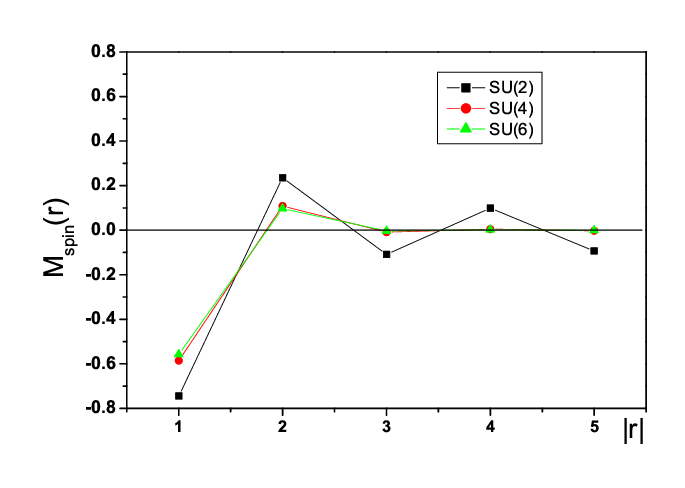}
\caption{ The normalized spin-spin correlations as functions of
distance (along the $x$-axis) with  $T/t=0.1$,  $U/t=4$ and
$2N=2,4$, and 6.} \label{fig:real}
\end{figure}

{\it Spin-spin correlations in real space} In Fig.\ref{fig:real}, we
plot the renormalized equal time spin-spin correlations for the
SU($2N$) Hubbard model as a function of distance defined as \bea
M_{spin}(r)=\frac{1}{(2N)^2-1} \langle\sum_{\alpha,\beta}
S_{\alpha\beta}(\mathbf{i})
S_{\beta\alpha}(\mathbf{i}+r\mathbf{e_x})\rangle, \eea which exhibit
a staggered antiferromagnetic structure. For the case of SU(4) and
SU(6), spin-spin correlation functions decay much more drastically
than that of SU(2). This agrees with the fact that the AF
correlations of the SU($2N$) Hubbard model are weaken with
increasing $2N$.

{\it The charge gap} The charge gap is defined as the energy cost to
add one particle in the ground state of the system composed of
$N$-particles. Assume that \bea
\hat{H}|\Psi_0^{N+1}\rangle&=&E_0^{N+1}|\Psi_n^{N+1}\rangle, \nn \\
\hat{H}|\Psi_0^{N}\rangle&=&E_0^{N}|\Psi_n^{N}\rangle, \eea where
$\hat H$ is the Hamiltonian of the grand canonical ensemble for the
SU($2N$) Hubbard model as Eq. (1) in the main text. (The chemical
potential $\mu$ is set $0$ in Eq. (1)). The charge gap is
$\Delta_c=E_0^{N+1}-E_0^N$. The onsite time-displaced Green's
function for $\tau>0$ reads
\begin{widetext}
\begin{eqnarray}
\nonumber G^>(\vec{r}=0,\tau)&=&\frac{1}{L^2}\sum_{i}G^>(\tau)_{ii}
=\frac{1}{L^2}\sum_i\langle
\Psi_0^N|e^{\tau\hat{H}}c_ie^{-\tau\hat{H}}c^\dag_i|\Psi_0^N\rangle.
\end{eqnarray}
By inserting the complete set
$I=\sum_n|\Psi_n^{N+1}\rangle\langle\Psi_n^{N+1}|$, the above
equation becomes
\begin{eqnarray}
G^>(0,\tau)
=\frac{1}{L^2}\sum_{i,n}e^{-\tau(E_n^{N+1}-E_0^N)}\langle
\Psi_0^N|c_i|\Psi_n^{N+1}\rangle \langle \Psi_n^{N+1}|
c^\dag_i|\Psi_0^N\rangle
=\frac{1}{L^2}\sum_{i,n}e^{-\tau(E_n^{N+1}-E_0^N)}|\langle
\Psi_0^N|c_i|\Psi_n^{N+1}\rangle|^2.
\end{eqnarray}
\end{widetext}
Therefore, at large $\tau$, we have $G^>(\vec{r}=0,\tau)\sim
e^{-\tau\Delta_c}$ which can be used to estimate the value of
$\Delta_c$ \cite{assaad1996}.

\end{appendix}

\end{document}